\documentclass[%
 reprint,
superscriptaddress,
 amsmath,amssymb,
 aps,
]{revtex4-1}

\usepackage{graphicx}
\usepackage{dcolumn}
\usepackage{bm}

\begin{document}

\title{Shift current bulk photovoltaic effect influenced by quasiparticles and excitons}

\author{Ruixiang Fei}
\affiliation{Department of Chemistry, University of Pennsylvania,
	Philadelphia, Pennsylvania 19104-6323, USA%
}%

\author{Liang Z. Tan}
\affiliation{Molecular Foundry, Lawrence Berkeley National Laboratory, Berkeley, California 94720, United States} 

\author{Andrew M. Rappe}
\email{rappe@sas.upenn.edu}
\affiliation{Department of Chemistry, University of Pennsylvania,
	Philadelphia, Pennsylvania 19104-6323, USA%
}%


\begin{abstract}
We compute the shift current bulk photovoltaic effect (BPVE) in bulk BaTiO$_3$ and two-dimensional monochalcogenide SnSe considering quasi-particle corrections and exciton effects. 
We explore changes in shift current peak position and magnitude reduction due to band renormalization. For BaTiO$_3$, we demonstrate that shift current is reduced near the band edge due to exciton effects. Comparison of these results with experiments on BaTiO$_3$ indicate that mechanisms other than shift current may be contributing to BPVE. Additionally, we reveal that the shift current near the band gap shows only a small change due to excitons in two-dimensional SnSe, suggesting that the thin film geometry provides a feasible way to reduce the exciton effect on the shift current.
These results suggest that many-body corrections are important for accurate assessments of bulk photovoltaic materials and to understand the mechanisms behind the BPVE.
\end{abstract}

\maketitle

The bulk photovoltaic effect (BPVE), which has also been referred to as the 
``photogalvanic effect", is a resonant nonlinear process where photocurrent is generated in the bulk of materials\cite{laman2005ultrafast, sipe2000Second}.
The BPVE requires a lack of 
inversion symmetry, allowing an asymmetric photoexcitation of carriers \cite{belinicher1980photogalvanic,sturman1992photovoltaic,glass1995high}. 
Because the photovoltage is not limited by the band-gap energy, and a p-n junction
or interface is not required, the BPVE in ferroelectrics has attracted 
a lot of attention \cite{spanier2016power,ji2010bulk,Zenkevich2014giant}. Ferroelectric oxide materials including BaTiO$_3$ \cite{Chynoweth1956,koch1975bulk,koch1976anomalous}, LiNbO$_3$
\cite{Glass1974,fridkin1978anomalous} and BiFeO$_3$ \cite{choi2009,yang2010,Seidel2011,Young2012BFO} have been widely studied for their photovoltaic properties, with substantial effort 
devoted to understanding their origins. Deeper understanding of photovoltaic effects is crucial for the discovery and the design of
new types of ferroelectric semiconductors, including organic and hybrid materials\cite{zheng14p31,liu17p6500}, topological materials \cite{Tan16p237402}, and layered two-dimensional materials \cite{cook17p14176} for BPVE applications. More than one mechanism could contribute to the DC photocurrent, including the shift current \cite{sipe2000Second,tan2016shift} and the ballistic current \cite{Belinicher1978, ivchenko1984, Belinicher1988}, with their relative magnitudes currently under debate. The shift current is the result of the movement of the center of charge during optical excitation, e.g., transitions from valence to conduction bands. In similar fashion, transitions from defect levels to conduction bands should give rise to shift currents as well. 
In this paper, we focus on the shift current, which has been a topic of current research, and show that many-body effects which are often neglected in its computation give rise to sizable corrections. 

Comparisons of the experimentally measured BPVE in ferroelectric BaTiO$_3$ \cite{koch1975bulk,koch1976anomalous} and BiFeO$_3$ \cite{wei2011} with first-principles DFT calculations\cite{young2012first,Young2012BFO} suggests that shift current is responsible for a significant portion of the BPVE in ferroelectrics. However, these conclusions were drawn from calculations neglecting quasiparticle corrections and excitonic effects, and should be revised with these many-body effects taken into account. Similarly, many theoretical predictions of shift current in new materials are routinely made without these many-body effects taken into account.

In the present Letter,
we study the shift current with quasiparticle GW corrections
and excitonic effects in the typical perovskite oxide BaTiO$_3$ and  large BPVE two-dimensional monochalcogenide SnSe \cite{Fei2016Ferro, Neaton2017RPL, Bellaiche2017PRL}. These many-body effects work to redistribute the spectral weight of the shift current response. In general, there is a tendency for the shift-current response to be reduced by the exciton effects, although the behaviors are different for bulk material and two-dimensional materials, as we show below.


In Refs. \cite{sipe2000Second, young2012first,1981BaltzTheory}, shift current is calculated within perturbation theory, with the monochromatic electric field treated classically, taking the form $E_s(t) = E_s(\omega)e^{i\omega t} + E_s(-\omega)e^{-i\omega t}$. The second order response function for the shift current includes transitions of electrons to all unoccupied bands, 

\begin{equation} 
\begin{split}
\label{shift_current}
j_{Q}(\omega) =& \sum_{s} \sigma_{ssQ}(0, \omega, -\omega)E_s(\omega)E_s(-\omega)\\
\sigma_{ssQ}(0, \omega, -\omega) = &\pi\frac{e^3}{\hbar^2}\int{\frac{d\textbf{k}}{4\pi^3}}\sum_{nm} f_{mn} r_{s}(m,n,\textbf{k})\\
&\times r_{s}(n, m,\textbf{k})R_{Q}(m, n, \textbf{k})\delta(\omega_{mn}\pm\omega)
\end{split}
\end{equation}
where $n$ and $m$ are the band indices,  $\textbf{k}$ is the wave vector, $f_{mn}=f_m-f_n$ is the Fermi-Dirac occupation number, $\omega_{mn}=\omega_m-\omega_n$ is the band energy difference and $\sigma_{ssQ}$
is a third-rank tensor giving current density $J$ as a response to  monochromatic electromagnetic
field $E$. 

The expression is composed of the effective position matrix elements $r_{s}(m, n, \textbf{k})$ and 
the so-called ``shift vector" $R_{Q}(m, n, \textbf{k})$:
\begin{equation} 
\begin{split}
\label{matrix_shift}
r_s(m, n, \textbf{k})\equiv & \frac{v_s(m, n, \textbf{k})}{i\omega_{mn}} = \frac{\langle m\textbf{k}|v_s|n\textbf{k}\rangle}{i\omega_{mn}}\\
R_{Q}(m, n, \textbf{k})= &-\frac{\partial\phi(m,n, \textbf{k})}{\partial k_Q}-(A_Q(n,n,\textbf{k})-A_Q(m,m,\textbf{k}))
\end{split}
\end{equation}
Here, $v(m, n, \textbf{k})$ are velocity matrix elements, $A(m,m,\textbf{k})$ are Berry connections for band m, and $\phi(m,n, \textbf{k})$ is the phase of the momentum matrix element between bands $m$ and $n$.

The wave functions and eigenvalues were generated using the plane-wave density functional
theory (DFT) package Quantum ESPRESSO with the generalized gradient approximation (GGA) exchange
correlation functional. Norm-conserving, designed nonlocal pseudopotentials \cite{1990Rappe,1999Ramer} were produced using the OPIUM package. Quasiparticle corrections to the nonlinear conductivity $\sigma$ were made by using GW-renormalized matrix elements~\cite{Ibanez-azpiroz18p245143} and quasiparticle energies in Eq.~\ref{shift_current} (See Eq.~S6 for details).
The current density induced by an external light source depends not only on the nonlinear conductivity, but also on the attenuation of the light field within the material. To account for this effect, we consider corrections to the dielectric function at the  GW+Bethe Salpeter Equation (BSE) level. The BSE

\begin{equation} 
\begin{split}
\label{BSE_equation}
(E_{c\textbf{k}}-E_{v\textbf{k}})A_{vc\textbf{k}}^S+ 
\sum_{v'c'\textbf{k}'} K_{vc\textbf{k}, v'c'\textbf{k}'}(\Omega^S)A_{v'c'\textbf{k}'}^S
=\Omega^S A_{vc\textbf{k}}^S
\end{split}
\end{equation}
gives correlated e-h excitations $S$ of energy $\Omega^S$, expanded
in the basis of e-h pairs $|S\rangle = \sum A_{vc\textbf{k}}^S|vc\textbf{k}\rangle$. Here, $v$ and $c$ stand for the valence and conduction band indices, respectively. $K$ is e-h interaction kernel. 
The quasiparticle and excitonic effects are incorporated into shift-current calculation by interfacing our
in-house shift current code \cite{young2012first, Young2012BFO} with the BerkeleyGW package \cite{1986Hybertsen, 2000Rohlfing}. For additional information, see the Supplemental Materials \cite{Supplement}.

\begin{figure}
	\centering
	\includegraphics[scale=0.28]{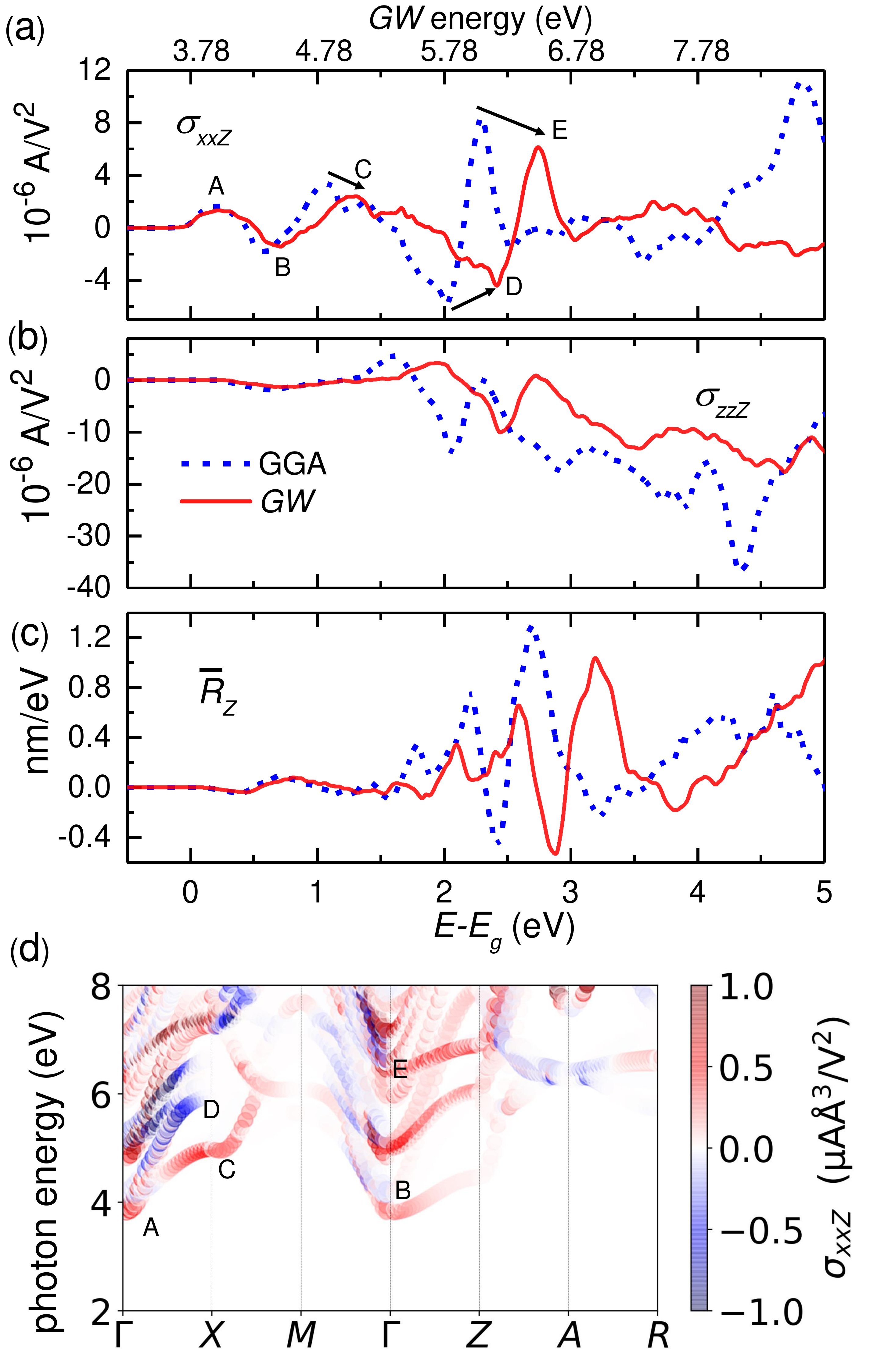}
	\caption{ The overall current susceptibility $\sigma_{xxZ}$ (a), $\sigma_{zzZ}$ (b) and aggregated 
		shift vector $\bar{R}_{Z}$ (c) for BaTiO$_3$ as a function of energy above
		their respective band gaps.	In each panel of (a-c), spectra are given at the GGA level (blue) and the GW level (red); the corresponding excitation energy of spectra at the GW level are given on the top. In (a), the arrows represent the corresponding peaks between GGA and GW levels. 
		The $k$-resolved photocurrent $\sigma_{xxZ}$ at the GW level (d), in which the color gives the value of the photocurrent response. The dominant contributions to the photocurrent peaks at the GW level are labeled in (d), corresponding to labeled peak in (a). The calculations adopt the experimental lattice constants at room temperature $a=b=3.9998$ \AA, and $c=4.018$ \AA
	}
	\label{Fig_1} 
\end{figure}

We first perform first-principles DFT calculations of the shift current for BaTiO$_3$ (BTO), 
which derives from the cubic perovskite structure and is in a tetragonal ferroelectric phase 
at room temperature. We use experimental room temperature geometries \cite{buttner1992} for this calculation. The quasiparticle energies are calculated with the $G_0W_0$ approximation. Because the quasiparticle wavefunction is equal to the DFT wavefunctions at the first order, we do not update the wavefunctions for shift current tensor calculation at the GW level.  In Fig 1 (a-c), we show the shift current tensor elements in the direction of material polarization ($Z$), and the shift vector integrated over the Brillouin zone \={R} given by

\begin{equation} 
\label{shift_vector}
\begin{split}
\bar{R}_Z(\omega) &= \sum_{nm}\int d\textbf{k} R_z(m,n,\textbf{k})\delta(\omega_{m}-\omega_{n}\pm\omega)
\end{split}
\end{equation}


The direct band gap at the GGA level is 2.10 eV, and the quasiparticle direct band gap is 3.78 eV, consistent with previous GW calculations \cite{Sanna2011}. To compare spectra at both GGA and GW levels, we plot the spectra as a function of energy above their respective band gaps. The peak of response is several eV above the band gap and well outside the visible spectrum, while the shift current at energies near the band gap is small. 

GW corrections in general increase band gaps and bandwidths \cite{hybertsen85p1418} in semiconductors. We therefore expect that the effect of GW corrections on a shift current spectrum is to shift and stretch the spectrum to higher frequencies. Peak position changes, indicated by arrows in Fig 1(a), are stronger at the high energy, but tiny at low energy (e.g. peak A). The corresponding $k$-space-resolved photocurrent at the GW level is illustrated in Fig 1(d), with colors representing the current direction. 
Besides the changes in the spectral peak position at the GW level, the magnitudes of the spectra after GW correction are smaller. Because of the bandwidth increase, the stretching of the spectrum distributes spectral weight over a greater spectral range, resulting in lower magnitudes of shift current. The GW increases the valence and conduction bandwidths by $22\%$ and $20\%$ for BTO, respectively, showing that the quasiparticle correction in BTO is important for accurate shift current spectra.




\begin{figure}
	\centering
	\includegraphics[scale=0.17]{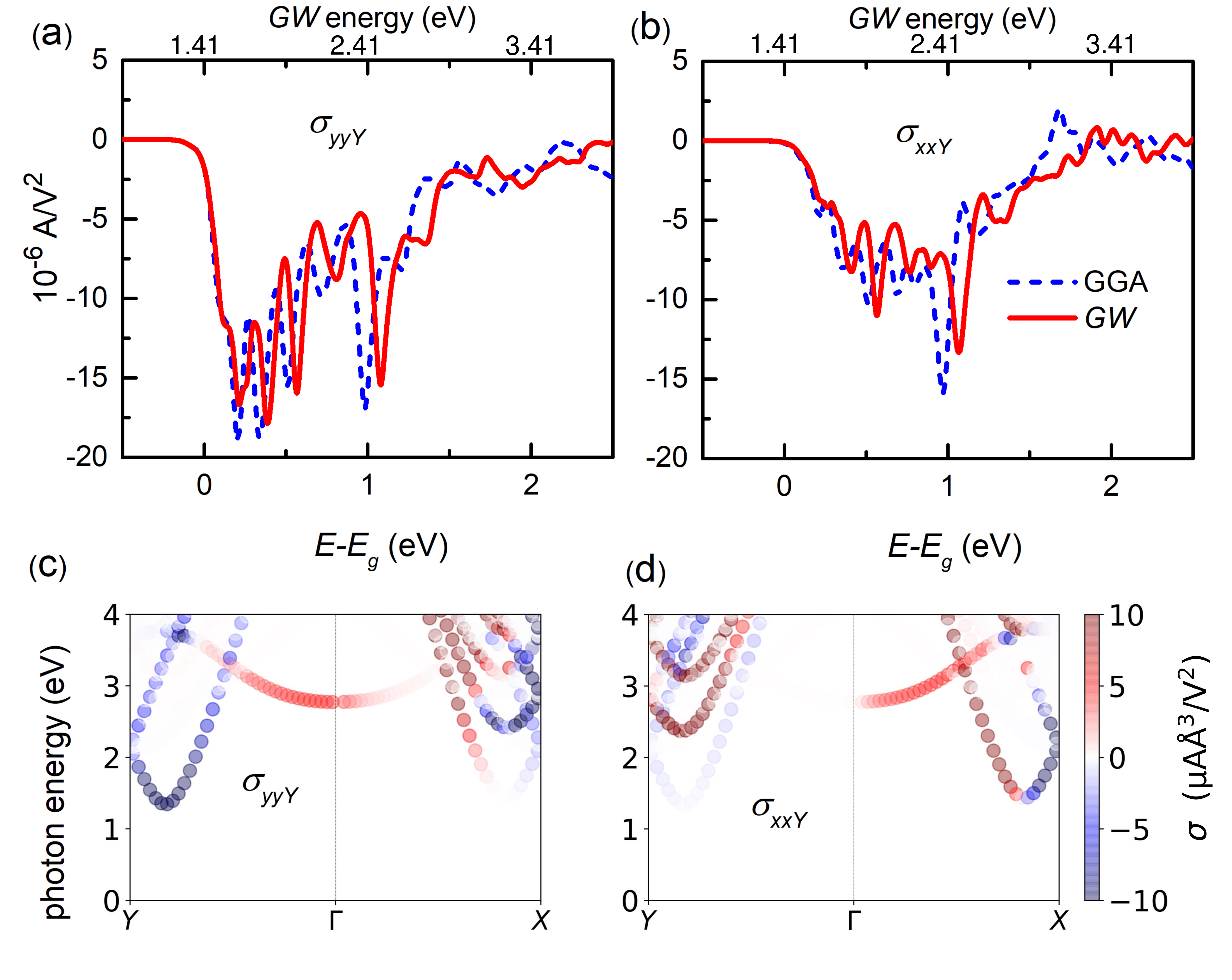}
	\caption{The current susceptibility $\sigma_{yyY}$ (a) and $\sigma_{xxY}$ (b) for mono layer SnSe as a function of energy above their respective band gaps. In (a) and (b), spectra are given at the GGA level (blue) and GW level (red), and the correspond energy of spectra at the GW level are given on the top of each panel. The $k$-resolved photocurrent susceptibility $\sigma_{yyY}$ (c) and $\sigma_{xxY}$ (d) at the GW level. In (c) and (d), the color gives the value of the photocurrent response, and each direction valley in Brillouin zone can be optically pumped separately by excitation with linearly polarized light.
	}
	\label{Fig_2} 
\end{figure}

Next, we apply this GW shift current analysis to the monolayer monochalcogenides, a class of room-temperature two-dimensional ferroelectrics\cite{Fei2016Ferro, chang2016p6296}. We select  SnSe as a prototype, which has large spontaneous polarization $0.3$ C/m$^2$ in the $Y$ direction \cite{Fei2016Ferro,Neaton2017RPL} and large shift current susceptibility \cite{Neaton2017RPL}. The GGA and GW band gaps of 0.92 eV and 1.41 eV are in the optimal range for solar cells \cite{franzman2010p4060,shi2015p6926}.
There are large shift current responses under $yy$ (Fig 2a) and $xx$ (Fig 2b) polarized illumination within 1 eV above the band gap. Similar to BTO, We find that the GW shift current spectrum is shifted and stretched to higher frequencies, and the magnitudes after GW correction are smaller. However, the shift current response correction is weaker than that of bulk BTO. We attribute this to the smaller bandwidth correction for monolayer SnSe due to the stronger Coulomb screening, as the GW valence and conduction bandwidths increase by only $13\%$ and $8\%$, respectively. The shift current  susceptibility $\sigma_{xxY}$ near band gap is much smaller than $\sigma_{yyY}$. It can be understood from the $k$-resolved current susceptibility (Fig 2c and 2d). The $yy$ polarized light pumps more current for the $y$ valley. Each valley in Brillouin zone can be separately pumped  with linearly polarized light. 
Interestingly, different from the valley separated by circular-polarized light in MoS$_2$ \cite{zeng2012p490,xiao2012196802}, here it is separated by linear-polarized light.

Generally, the bandwidth of most bulk semiconductors is underestimated by more than 10\% at DFT level, while GW gives a more accurate bandwidth \cite{hybertsen85p1418,ishii2010p2150}. So the quasiparticle correction is suggested for shift current especially for the high-energy photon. On the other hand, the quasiparticle correction for shift-current of two-dimensional materials to be strong is not guaranteed, suggesting a case-by-case analysis of the quasipaticle correction on shift current.




\begin{figure}
	\centering
	\includegraphics[scale=0.20]{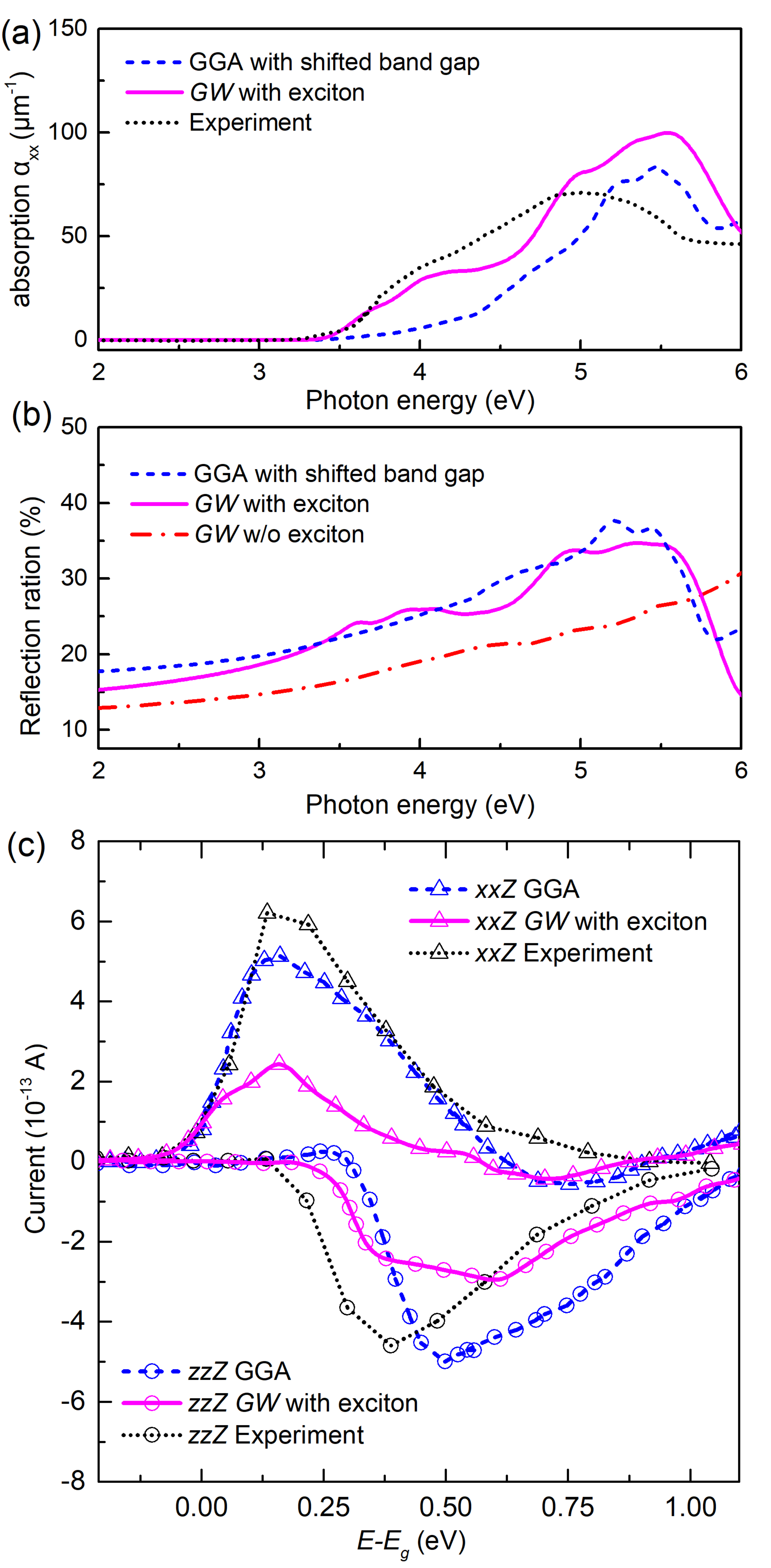}
	\caption{For BaTiO$_3$, (a) the experimental \cite{cardona1965p651} and computed absorption spectra as a function of photon energy. (b) The reflection ration computed with GGA, GW with exciton, and GW without exciton. The GGA absorption spectrum and reflection ratio are shifted to the experiment band gap. (c) The experimental current \cite{koch1975bulk} and  GGA current and GW with exciton correction current, for transverse ($xxZ$) and longitudinal ($zzZ$) electric field orientation, as a function of energy above their respective band gaps. The solid and dashed lines are calculated results for a choice of experimental parameters of 0.5 mW/cm$^2$ illumination intensity and 0.15 cm sample width. }


	\label{Fig_3} 
\end{figure}

Next, we analyze the impact of the excitons on absorption and reflectivity, and their effect on shift current.
The total current in the direction normal to light incidence is 
\begin{equation} 
\begin{split}
\label{current}
J_{ssQ} 
        &=\frac{K_{ssQ}}{\alpha_{ss}(\omega)}(1-R_{ss})(1-e^{-\alpha_{ss}(\omega)d})wI_{s}\\
K_{ssQ} &=\frac{2\sigma_{ssQ}}{c\epsilon_0 \sqrt{\epsilon_r}}
\end{split}
\end{equation}
where $\alpha_{ss}$ is the absorption coefficient, $R_{ss}$ is reflection ratio, $I_s$ is the
light intensity for light polarization direction $s$, 
$w$ is the width of the crystal surface exposed to illumination, $d$ is depth of the crystal, and $\epsilon_r$ is dielectric constant. For a bulk crystal, e.g. BaTiO$_3$, the depth $d$ is much larger than length scale of inverse absorption coefficient $\alpha_{ss}^{-1}$ (hundreds of \AA), and the total current is
\begin{equation} 
\begin{split}
\label{current}
J_{ssQ} &=\frac{\sigma_{ssQ}}{\alpha_{ss}(\omega)}(1-R_{ss})wE_{s}^2=G_{ssQ}(1-R_{ss})wI_{s}\\
\end{split}
\end{equation}
where $G_{ssQ}$ is the Glass coefficient \cite{Glass1974}. 
The absorption coefficient is calculated using $\alpha(\omega) = \frac{\omega}{c} \sqrt{2\sqrt{\epsilon_1^2+\epsilon_2^2}-2\epsilon_1}$, where $\epsilon_1$ and $\epsilon_2$ are obtained from the BSE calculations. See Supplemental Material for more details \cite{Supplement, giuliani2005, dresselhaus2018}. 


From Eq.(6), we see that the shift current is highly dependent on the absorption coefficient for bulk materials. 
For bulk single-crystal BTO, the experimental absorption coefficient from Ref.~\cite{cardona1965p651} is compared to the coefficients computed using GGA and GW with exciton effects, shown in Fig 3a. Even with a shift of the GGA absorption spectrum to the experimental band gap, it is still qualitatively incorrect compared to experiment. The absorption coefficient is highly underestimated by the GGA calculation, while the GW with exciton correction gives much better absorption coefficient. 
The enhancement of $\epsilon_2$ near the band edge induced by exciton significantly influences absorption; the $\alpha(\omega)$ with exciton effects is larger than without, for photon energies within 1 eV above the band gap. 
Besides the absorption coefficient, the reflection ratio (or reflectivity), calculated from dielectric constant, is also influenced by excitons (see supplemental information). For the reflection ratio in BTO, experimental results suitable for quantitative comparison could not be located. However, the reflection ratio measured using unpolarized light and an unpoled sample is around 22\%-30\% \cite{cardona1965p651} within 1 eV above band gap, which is agree well with our GW calculation with exciton (Fig 3b). We also shift the GGA reflectivity to the experimental band gap. The GGA reflectivity is not very different from the GW with exciton correction. So the shift current difference between GGA and GW with exciton correction in BTO is dominated by the absorption coefficient correction.
In Fig 3c, the experimental current response from \cite{koch1975bulk} is compared to the shift current computed using GGA and GW with exciton correction, using the light intensity $0.5$ mW/cm$^2$ and crystal dimensions $0.15$ cm of the experiment\cite{koch1975bulk,koch1976anomalous}. The GW with exciton correction improves the energy alignment of transvers current ($xxZ$) response. However, the magnitude of the current including exciton effect is around half of the experimental value, suggesting that mechanisms other than the shift current may contribute to the BPVE. 


\begin{figure}
	\centering
	\includegraphics[scale=0.26]{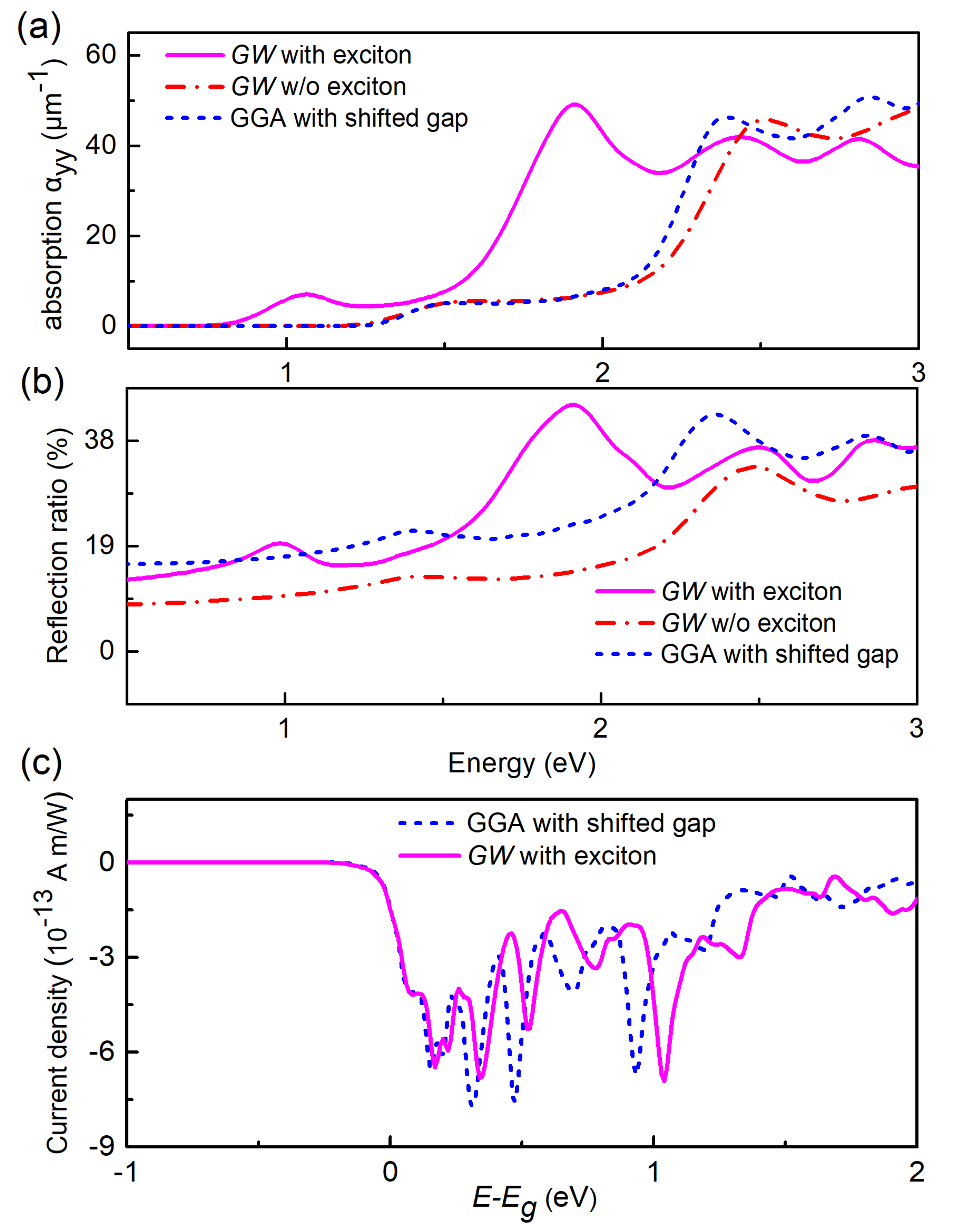}
	\caption{For two dimensional SnSe, (a) the absorption coefficient is computed at the GGA level, GW level, and GW with exciton correction and (b) reflection ratio, for longitudinal ($yy$) electric field orientation, as a function of photon energy. (c) the shift current density, for longitudinal ($yy$) electric field orientation, at GGA level and GW with exciton correction as a function of energy above the GW band gap. For comparison, we estimate the thickness of SnSe to be $5.5$ \AA. The GGA spectra in each panel are shifted to the GW band gap.
	}
	\label{Fig_4} 
\end{figure}

Next, we consider the excitonic effects in two-dimensional semiconductor SnSe. We use Coulomb slab truncation and 35 \AA vacuum in the GW and exciton calculations, which was sufficient for convergence (see Supplemental Material). Fig 4(a) and (b) show the absorption coefficient and reflection ratio at different levels of calculation, with similar features as for BTO. The absorption coefficient with excitonic effects is enhanced near the band gap, and the reflection ratio including GW with exciton is renormalized, compared with GGA calculations. Nevertheless, these renormalizations do not have a strong effect on the photocurrent of two-dimensional SnSe because its thickness $d$ ($5.5$ \AA) is much smaller than the inverse absorption coefficient, e.g. $\alpha_{yy}^{-1}=2000$ \AA 
for 1.41 eV light.
So, Eq.(5) is reduced to $J_{ssQ}=K_{ssQ}(1-R_{ss})dwI_s$ for two-dimensional materials. As a result, excitonic effects on the absorption coefficient has a weak influence on the shift current of two-dimensional materials. The overall shift current density (Fig 4c) with exciton is renormalized and slightly reduced from the GGA calculated shift current, caused by the renormalized reflectivity due to excitons and band stretching due to quasiparticles. 
Importantly, the small reduction in the magnitude of the shift current near the band edge implies that the exciton effect plays much less role in this case and the reason is that the optical penetration depth is much larger than the thickness of the material.

In summary, we have demonstrated that many-body effects lead to significant changes in the shift current response. Quasiparticle GW corrections lead to shifts in peak position and reductions in shift current magnitude, while excitonic effects on the absorption and reflection result in rearrangements of spectral weight, reducing shift current response near the band gap. These results have consequences for our understanding of the role of shift current in the BPVE, and our assessments of the performance of shift current materials. Comparisons of our BaTiO$_3$ calculations with experiment indicate that other mechanisms are likely to play a role in the BPVE. The two-dimensional SnSe calculations reveal that excitons have very small influence near the band gap, suggesting in-plane two-dimensional or thin film geometries can optimize the materials for BPVE application.



R. F. thank Wenjie Dou and Shiyuan Gao for valuable discussions.
This work was supported by the U.S. Department of  Energy, Office of Basic Energy Sciences, under Grant No. DE-FG02-07ER46431. The authors acknowledge computational support from the NERSC of the DOE.

\bibliography{exciton}
\end{document}